# SYMMETRY ENERGY FOR NUCLEI BEYOND THE STABILITY VALLEY

V.M. Kolomietz and A.I.Sanzhur

*Institute for Nuclear Research, 03680 Kiev, Ukraine*

We apply the direct variational method to derive the equation of state for finite nuclei within the stability valley. The extended Thomas-Fermi approximation for the energy functional with Skyrme forces is used. Using the leptodermous expansion for the profile nucleon densities, we have studied the neutron coat and the isospin symmetry energy for neutron rich nuclei. Using equation of state for the pressure, we derive the region of spinodal instability of finite nuclei and its dependence on the mass number, the asymmetry parameter and the Skyrme force parameters. We suggest the procedure of derivation of the isospin symmetry energy from analysis of the isospin shift of chemical potential $\Delta\lambda = \lambda_n - \lambda_p$ beyond the beta-stability line.

## 1. Introduction

The isospin symmetry energy (ISE) is one of the key elements of the nuclear equation of state. Many static and dynamic features of nuclei are sensitive to isospin degree of freedom. The expansion of the nuclear stability valley as well as the existence of the isovector eigenvibrations depend significantly on the ISE [1,2]. Solution of some problems of nuclear collisions such as the isoscaling effect, the nuclear multifragmentation accompanying by the emission of asymmetric clusters, the isospin instability of nuclei at high temperatures, etc. depends on our knowledge of the nuclear equation of state for the ISE, namely, its dependence on the particle density and the asymmetry parameter $X = (N-Z)/A$. The related problem is the investigation of the nucleon redistribution within the surface region of the nucleus, in particular, the "neutron" coat and the neutron excess for the nuclei far away from the beta stability line.

## 2. Energy functional for asymmetric nuclei

We will follow the extended Thomas-Fermi approximation (ETFA) using the Skyrme-type force as the effective nucleon-nucleon interaction. In general, the total energy $E$ of the nucleus is the functional of the nucleon densities $\rho_p(\mathbf{r})$ and $\rho_n(\mathbf{r})$

$$E = \int d\mathbf{r}\ \varepsilon_{tot}(\mathbf{r}) \equiv \int d\mathbf{r}\ \varepsilon_{tot}[\rho_n(\mathbf{r}), \rho_p(\mathbf{r})] \qquad (1)$$

and depends on the particle density and its gradients only. We point out that the ETFA is the one possible realization of general Hohenberg-Kohn theorem in many body problem [3]. The unknown values of $\rho_n$ and $\rho_p$ can be evaluated from the condition of equilibrium. The equilibrium condition can be written as a Lagrange variational problem. Namely,

$$\delta(E - \lambda_n N - \lambda_p Z) = 0, \qquad (2)$$

where the variation with respect to all possible small changes of $\rho_n$ and $\rho_p$ is assumed. The Lagrange multipliers $\lambda_n$ and $\lambda_p$ are the chemical potentials of the neutrons and the protons, respectively, which are fixed by the condition that the number of particles is conserved

$$N = \int d\mathbf{r}\, \rho_n(\mathbf{r}), \qquad Z = \int d\mathbf{r}\, \rho_p(\mathbf{r}). \tag{3}$$

The total energy density $\varepsilon_{tot}[\rho_n, \rho_p]$ in Eq. (1) includes the kinetic energy density, $\varepsilon_{kin}[\rho_n, \rho_p]$, the potential energy of $NN$-interaction, $\varepsilon_{pot}[\rho_n, \rho_p]$, and the Coulomb energy, $\varepsilon_C[\rho_p]$,

$$\varepsilon_{tot}[\rho_n, \rho_p] = \varepsilon_{kin}[\rho_n, \rho_p] + \varepsilon_{pot}[\rho_n, \rho_p] + \varepsilon_C[\rho_p]. \tag{4}$$

In the framework of ETFA, both $\varepsilon_{kin}[\rho_n, \rho_p]$ and $\varepsilon_{pot}[\rho_n, \rho_p]$ depend on the nucleon densities $\rho_n$ and $\rho_p$, and its gradients. Their explicit form is given in Refs. [4] and [5].

Considering the asymmetric nuclei with $X = (N-Z)/A \ll 1$, we will introduce the new trial densities, namely the total density $\rho_+ = \rho_n + \rho_p$ and the neutron excess density $\rho_- = \rho_n - \rho_p$. In this paper, we will not solve the Euler-Lagrange equation (2) for self-consistent densities $\rho_q$ ($q = n, p$) or $\rho_\pm$. Instead, we will follow the direct variational method (DVM) and assume the density profile function $\rho_\pm$ to be given by a power of the Fermi function as follows

$$\rho_+(r) = \rho_{+,0} f(r), \quad \rho_-(r) = \rho_{-,0} f(r) - \frac{1}{2} \rho_{+,0} \Delta \frac{df(r)}{dr}. \tag{5}$$

Here,

$$f(r) = \left[1 + \exp[(r-R)/a]\right]^{-\delta}, \tag{6}$$

the values $\rho_{+,0}$ and $\rho_{-,0}$ are related to the bulk density, $R$ is the nuclear radius, $a$ is the diffuseness parameter and $\Delta$ is the parameter of neutron skin. Parameter $\delta$ in Eq. (6) determines the behavior of the trial functions in the nuclear surface region. The unknown variational parameters $\rho_{\pm,0}, R, a, \Delta$ and $\delta$ must be derived from the variational principle (2). The number of variational parameters for the trial functions (5) is reduced due to the restrictions of Eq. (3). Below we will assume the leptodermous condition $a/R \ll 1$ which reduce the variational problem to four independent parameters $\rho_{+,0}, a, \Delta$ and $\delta$. We point out that significant advantage of the used DVM is the possibility to establish the equation of state for *finite* nuclei, i.e., dependence of the energy per nucleon on the mass number $A$, the asymmetry parameter $X$ and the bulk density $\rho_{+,0}$.

The parameter $\Delta$ is related to the number, $N_\Delta$, of neutrons in surface region of the nucleus ("neutron coat"). Substituting Eqs. (5) and (6) into Eq. (3) and using the leptodermous expansion, we obtain for the neutron excess $N - Z$ the following expression

$$N - Z \approx \frac{4\pi}{3} R^3 \rho_{-,0} + 4\pi R^2 \frac{\rho_{+,0}}{2} \Delta. \tag{7}$$

The first term $\sim R^3$ on the right hand side of Eq. (7) is due to redistribution of the neutron excess within the nuclear volume

$$N_V \approx \frac{\rho_{-,0}}{\rho_{+,0}} A \tag{8}$$

while the second term $\sim R^2$ is the number of neutrons within neutron coat

$$N_\Delta \approx \frac{3}{2} \left( \frac{4\pi}{3} \rho_{+,0} \right)^{1/3} A^{2/3} \Delta. \tag{9}$$

Due to the leptodermous condition, the total energy (1) takes the following form of $A, X$-expansion

$$E/A \equiv e_A = e_0(A) + e_1(A)X + e_2(A)X^2 + E_C(X)/A, \tag{10}$$

where

$$e_i(A) = c_{i,0} + c_{i,1} A^{-1/3} + c_{i,2} A^{-2/3} \tag{11}$$

and $E_C(X)$ is the total Coulomb energy.

Note that we have omitted in Eq. (11) the term $\sim A^{-1}$ which leads to the $A$-independent shift in the total energy $E$. The coefficients $c_{i,j}$ in Eq. (11) are the functions of the parameters $\rho_{+,0}, a, \Delta$ and $\delta$. The last term $\sim X^2$ in Eq. (10) gives rise to the isospin symmetry energy for stable nuclei as well as for nuclei beyond the beta-stability line. For the fixed values of $A$ and $X$, the basic probe parameters $\rho_{+,0}, a, \Delta$ and $\delta$ are evaluated from the variational conditions

$$\frac{\partial E/A}{\partial p_\nu} = 0, \qquad \nu = 1,...,4, \tag{12}$$

where $E/A$ is taken from Eq. (10) and we use the following short notations for the variational parameters $\{p_\nu\} = \{\rho_{+,0}, a, \Delta, \delta\}, \quad \nu = 1,...,4.$

The beta-stability line $X = X^*(A)$ can be directly derived from Eq. (10) by the condition

$$\left. \frac{\partial E/A}{\partial X} \right|_A = 0 \quad \Rightarrow \quad X^*(A) = -\frac{e_1(A) - 2e_C(A)}{2[e_2(A) + e_C(A)]}, \tag{13}$$

where

$$e_C(A) = \frac{3}{20} e^2 \left( \frac{4\pi \rho_{+,0}}{3} \right)^{1/3} A^{2/3}$$

Along the beta-stability line, the binding energy per particle is then given by

$$E^*/A = e_0^*(A) + e_1^*(A)X^* + e_2^*(A)X^{*2} + E_C(X^*)/A, \tag{14}$$

where the upper index "$*$" indicates that the corresponding quantity is determined by the variational conditions (12) taken for fixed $A$ and $X = X^*$. Using Eq. (13) for a fixed value of $A$, the binding energy can be written beyond the beta-stability line as

$$E/A = E^*/A + b_I(A, X)(X - X^*)^2 + \Delta E_C(X)/A, \tag{15}$$

where $\Delta E_C(X)/A \approx e_C^*(A)(X-X^*)^2$ and

$$b_I(A,X) = b_{I,vol}(X) + b_{I,surf}(X)A^{-1/3} + b_{I,curv}(X)A^{-2/3}. \tag{16}$$

An additional $X$-dependence in $b_I(A,X)$ in Eqs. (15) and (16) occurs because, for each fixed $X$ in a small vicinity of $X^*$, we have to solve the variational equations (12) under the additional conditions of Eq. (3) and, consequently, the variational parameters $\rho_{+,0}, a, \Delta$ and $\delta$ become $X$-dependent (see also Fig. 2). Using Eq. (15), one can establish the important relation for the chemical potential $\lambda_q$ beyond the beta-stability line. Namely, for fixed $A$, we obtain from Eqs. (15) and (13) the following result

$$\Delta\lambda = \lambda_n - \lambda_p = \left.\frac{\partial E}{\partial N}\right|_Z - \left.\frac{\partial E}{\partial Z}\right|_N = \left.\frac{2}{A}\frac{\partial^2 E}{\partial X^2}\right|_{A,X^*}(X-X^*) \approx 4\left[b_I(A,X^*) + e_C^*(A)\right](X-X^*). \tag{17}$$

On the beta-stability line, one has from Eq. (17) that $\Delta\lambda = 0$, as it has to be from the definition of the beta-stability line.

### 3. Numerical calculations

For the fixed mass number $A$ and asymmetry parameter $X$, the variation procedure in Eq. (2) with respect to all possible changes of parameters $\rho_{+,0}, a, \Delta$ and $\delta$ allows us to derive the parameters $R$ and $\rho_{-,0}$ and then the particle density $\rho_\pm$ of Eq. (5) as well as the binding energy $E/A$ of Eq. (15). We have performed calculations using the SkM, SLy230b and SIII force [5,6,7]. In Fig. 1 we compare the results for the beta stability line $Z = Z^*(N)$ obtained from Eq. (13) with the experimental data.

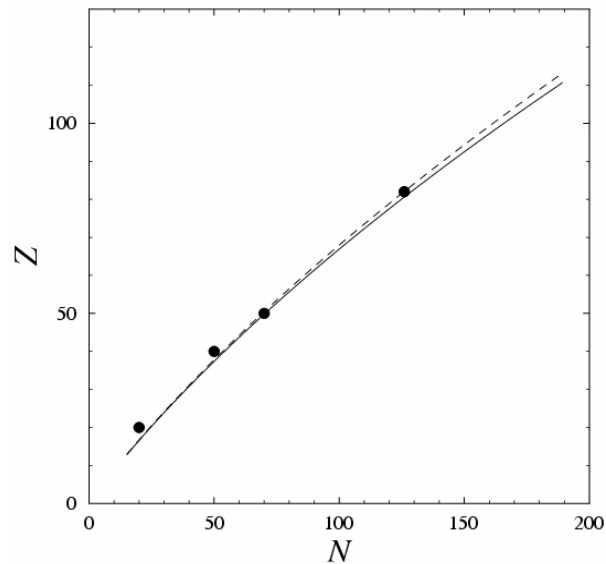

Fig.1. Line of beta stability $Z = Z^*(N)$. Solid and dashed lines are obtained from Eq. (13) for SkM and SLy230b sets of Skyrme force parameters; dots are the experimental data.

The results of calculations of symmetry energy parameters $b_{I,vol}$, $b_{I,surf}$ and $b_{I,curv}$ in Eq. (16) are shown in Table 1.

| Force | $X^*$ | $b_{I,vol}$, MeV | $b_{I,surf}$, MeV | $b_{I,curv}$, MeV |
|---|---|---|---|---|
| SkM | 0.221 | 30 | -53 | 63 |
| SLy230b | 0.211 | 32 | -50 | 43 |
| SIII | 0.213 | 29 | -31 | 10 |

Table 1. The results of the ETF calculation of the equilibrium asymmetry parameter $X^*$ and the isospin symmetry energies $b_{I,vol}$, $b_{I,surf}$ and $b_{I,curv}$ in Eq. (16) for the nucleus with mass number $A = 208$ for three sets of Skyrme forces SIII, SkM and SLy230b from Refs. [5,6,7].

Table 1 gives the coefficients in the vicinity of the beta-stability line. For the fixed mass number $A$, the values of $b_{I,vol}$, $b_{I,surf}$ and $b_{I,curv}$ in Eq. (16) are slightly dependent on the asymmetry parameter $X$. This effect is illustrated in Fig. 2 for $A = 208$.

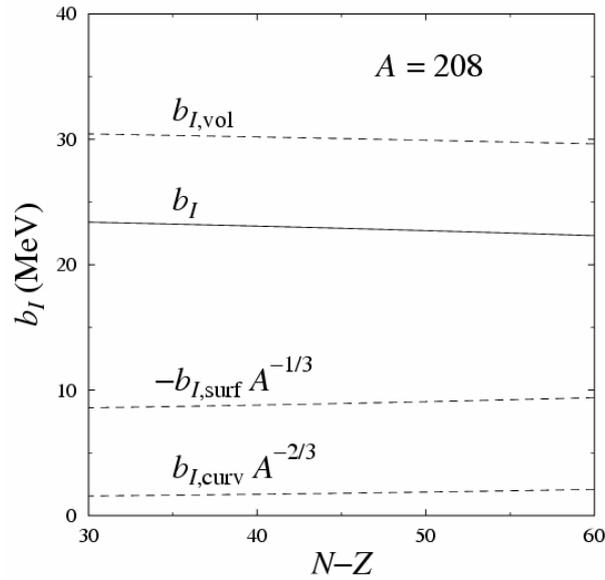

Fig. 2. Dependence of values $b_{I,vol}$, $b_{I,surf}$, $b_{I,curv}$ and $b_I$ Eq. (16) on the neutron excess $N - Z$ for the nuclei with fixed mass number $A = 208$. The calculation was performed with SkM forces.

In Fig. 3 we have plotted the $X$-dependence of the isotopic shift of the chemical potential, $\Delta\lambda = \lambda_n - \lambda_p$, for fixed mass number $A = 208$. As it seen from Fig. 3, there is the obvious correlation between calculation and experimental data. Thus, the quantity $\Delta\lambda$ can be used for the experimental determination of the value of symmetry coefficient $b_I$ from Eq. (16).

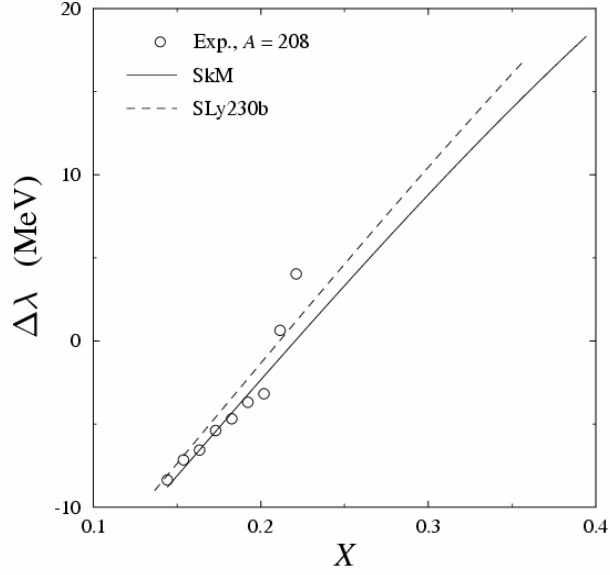

Fig. 3. Dependence of isotopic shift of the chemical potential $\Delta\lambda = \lambda_n - \lambda_p$ on the asymmetry parameter $X$ for fixed $A = 208$. Solid and dashed lines are obtained from Eq. (17) for SkM and SLy230b sets of Skyrme force parameters; dots are the experimental data from [8].

In general, the total energy (1) can be used to evaluate $E = E[\rho_+, \rho_-]$ beyond the equilibrium point $\rho_\pm = \rho_\pm^{(eq)}$. In particular, one can evaluate the equation of state, i.e., the dependence of the pressure $P$ on the bulk density $\rho_{+,0}$. We will derive the pressure $P$ as [9]

$$P = -\frac{\partial E}{\partial V}\bigg|_{A,X} = \rho_{+,0}^2 \frac{\partial E/A}{\partial \rho_{+,0}}\bigg|_{A,X}. \tag{18}$$

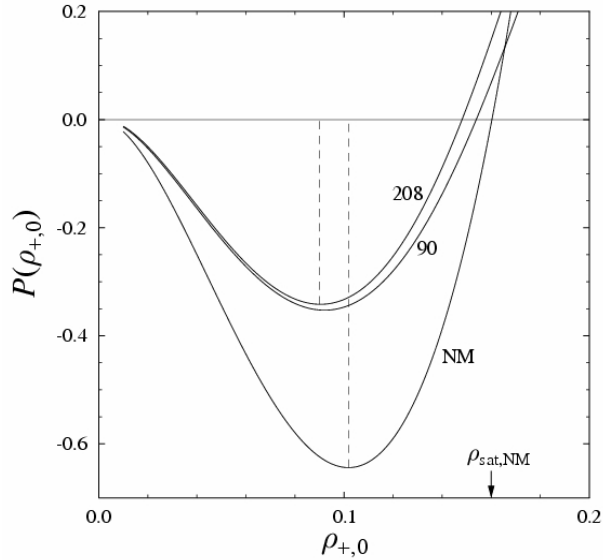

Fig. 4. The equation of state for the nuclei $^{90}$Zr, $^{208}$Pb and nuclear matter. The calculation performed for SkM forces. The area on the left hand side of dashed stright lines is the spinodal instability region for the nuclear matter and the nucleus $^{208}$Pb.

Note that the energy $E/A$ in Eq. (18) must be minimized with respect to $a, \Delta$ and $\delta$ for each fixed $\rho_{+,0}$. In Fig. 4 we have plotted the equation of state $P = P(\rho_{+,0})$ for two nuclei with $A = 90$ and $A = 208$. In agreement with Eq. (18), the equilibrium condition for the ground state of the nucleus at $\rho_{+,0} = \rho_{+,0}^{(eq)}$ means that $P(\rho_{+,0}^{(eq)}) = 0$. In Fig. 4, the minimum of the pressure $P(\rho_{+,0})$ is located at $\rho_{+,0} = \rho_{+,0}^{(crit)} \neq 0$. The nucleus is unstable within the spinodal instability region $\rho_{+,0} < \rho_{+,0}^{(crit)}$, where the incompressibility coefficient $K_A = 9 \partial P(\rho_{+,0}) / \partial \rho_{+,0}$ is negative, $K_A < 0$. In Table 2, we show the ratio $\rho_{+,0}^{(crit)} / \rho_{+,0}^{(eq)}$ for the nuclear matter and the finite nuclei for different Skyrme forces.

| Force | nucl. matter | $^{208}Pb$ | $^{120}Sn$ | $^{90}Zr$ |
|---|---|---|---|---|
| SkM | 0.635 | 0.610 | 0.602 | 0.598 |
| SLy230b | 0.639 | 0.614 | 0.605 | 0.601 |
| SIII | 0.674 | 0.657 | 0.651 | 0.648 |

Table 2. The results of the calculation of the ratio $\rho_{+,0}^{(crit)} / \rho_{+,0}^{(eq)}$ for three different forces used in Table 1.

A weak $A$-dependence of the critical density $\rho_{+,0}^{(crit)}$ and, consequently, of the spinodal instability region is mainly caused by both the surface tension and the Coulomb force which act in opposite directions. The interesting point is that the critical density $\rho_{+,0}^{(crit)}$ for the nuclear matter exceeds the one for finite nuclei. This is due to the fact that the gradient terms in Eq (4) give the effect on the surface and lead (without the Coulomb interaction) to an increase in the nucleon density in the center of the nucleus providing an additional stabilisation of finite nuclei with respect to the bulk density variation.

## 4. Summary and Conclusions

Starting from the energy functional of the extended Thomas-Fermi approximation and assuming the effective Skyrme-like forces, we have studied the ISE and the corresponding equation of state, namely, dependence of the ISE on the mass number $A$, the asymmetry parameter $X$ and the isoscalar bulk density $\rho_{+,0}$. Using the leptodermous expansion for the profile nucleon densities $\rho_p(\mathbf{r})$ and $\rho_n(\mathbf{r})$, we have established the $A$-dependence for the nuclear characteristics related to the isospin degree of freedom. A simple relation (7) has been obtained for the redistribution of the neutron excess $N - Z$ within the nucleus. An advantage of the direct variational method, used in this paper, is the possibility to derive the equation of state for finite nuclei, namely, dependence of the binding energy $E(\rho_{\pm,0})/A$, the pressure $P(\rho_{\pm,0})$, etc. on the bulk density $\rho_{\pm,0}$. The minimum of the pressure $P(\rho_{+,0})$ is located at $\rho_{+,0} = \rho_{+,0}^{(crit)} \neq 0$ which derives the region of the spinodal instability $\rho_{+,0} < \rho_{+,0}^{(crit)}$ for finite nuclei. We pointed out that the region of spinodal instability is slightly sensitive to the mass number $A$ and appears at the particle density $\rho_{+,0}^{(crit)} \approx 0.6\rho_{sat,NM}$, where $\rho_{sat,NM}$ is the saturation density for the nuclear matter. The critical density $\rho_{+,0}^{(crit)}$ is sensitive to the asymmetry parameter $X$ and the parameters of Skyrme forces. We have established the relation (17) between the isospin shift of chemical potential $\Delta\lambda = \lambda_n - \lambda_p$ and the isospin symmetry energy beyond the beta-stability line. This relation allowed us to evaluate the symmetry energy $b_I$ independently on the standard derivation from the mass formula. Moreover such kind of consideration performed for different mass number $A$, can be used to derive the "experimental" values of the $A$-dependent volume, surface and curvature terms in isospin symmetry energy. We have evaluated the contributions to the symmetry energy $b_I$ from the volume, $b_{I,vol}$, surface, $b_{I,surf}$, and curvature, $b_{I,curv}$, terms. Considering these values beyond the beta-stability line, see Fig. 2, we have noted that all of them dependent slightly on the asymmetry parameter $X$.

**References**

1. Ring P., Schuck P. The Nuclear Many-Body Problem. – Berlin: Springer-Verlag, 1980. – 718 p.
2. Myers W.D., Swiatecki W.J. Average Nuclear Properties // Ann. of. Phys. – 1969. – Vol. 55. – P. 395– 505.
3. Kohn W., Sham L. J. Self-Consistent Equations Including Exchange and Correlation Effects // Phys. Rev. – 1965. – Vol. 140. – P. A1133 – A1138.
4. Kolomietz V.M., Sanzhur A.I. Bulk and surface symmetry energy for the nuclei far from the valley of stability // Nuclear Physics and Atomic Energy – 2007. – No. 2 (20). – P. 7 – 15.
5. Brack M., Guet C., Håkansson H.-B. Selfconsistent semiclassical description of average nuclear properties – a link between microscopic and macroscopic models // Phys. Rep. – 1985. – Vol. 123. – P. 275 – 364.
6. Liu K.-F., Luo H., Ma Z., Shen Q. Skyrme-Landau parametrization of effective interactions (II). // Nucl. Phys. – 1991. – Vol. A534. – P. 25 – 47.
7. Chabanat E., Bonche P., Haensel P., Meyer J., Schaeffer R. A Skyrme parametrization from subnuclear to neutron star densities // Nucl. Phys. – 1997. – Vol. A627. – P. 710 – 746.
8. Martin M.J. Nuclear Data Sheets for $A = 208$ // Nucl. Data Sheets – 2007. – Vol. 108. – P. 1583 – 1806
9. Landau L.D., Lifshitz E.M. Statistical Physics – Oxford: Pergamon Press, 1958.